\def\year{2021}\relax
\documentclass[letterpaper]{article} 
\usepackage{aaai21}  
\usepackage{times}  
\usepackage{helvet} 
\usepackage{courier}  
\usepackage[hyphens]{url}  
\usepackage{graphicx} 
\def\UrlFont{\rm}  
\usepackage{natbib}  
\usepackage{caption} 
\usepackage{microtype}
\usepackage{booktabs} 
\usepackage[utf8]{inputenc}
\usepackage{url}
\usepackage{amsthm}
\usepackage{amssymb}
\usepackage{algorithmicx, algorithm}
\usepackage[noend]{algpseudocode}
\usepackage{verbatim}
\usepackage{graphicx}
\usepackage{amssymb}
\usepackage{tikz}
\usepackage{listings}
\usepackage{booktabs}
\usepackage{array}
\usepackage{chngcntr}
\usepackage{url}
\usepackage{mathtools}
\usepackage{xcolor}
\usepackage{nomencl}
\usepackage{siunitx,xinttools,xintfrac}
\usepackage{lipsum}
\usepackage{subfig}
\usepackage{xspace}
\usepackage{enumitem}
\usepackage{wrapfig}

\usepackage{pgfplots}
\usepackage{filecontents}
\usetikzlibrary{matrix}
\usepgfplotslibrary{groupplots}
\pgfplotsset{compat=newest}

\usepackage{hyperref}

\newcommand{\cita}[1]{\cite{#1}}

\theoremstyle{definition}
\newcommand{\pname}{\textbf{Minimum Robust Multi-Submodular Cover for Fairness}\xspace}
\newcommand{\pnametap}{Information Propagation for Multiple Groups\xspace}
\newcommand{\pnamemovie}{Movie Recommendation for Multiple Users\xspace}
\newcommand{\rfd}{\textsc{MinRF}\xspace}
\newcommand{\az}{\textsc{Alg0}\xspace}
\newcommand{\ao}{\textsc{Alg1}\xspace}
\newcommand{\ar}{\textsc{AlgR}\xspace}
\newcommand{\rg}{\textsc{RandGr}\xspace}
\newcommand{\sep}{\textsc{Sep}\xspace}
\newcommand{\djt}{\textsc{DisJoint}\xspace}
\newcommand{\gr}{\textsc{Greedy}\xspace}
\newcommand{\thr}{\textsc{ThresGr}\xspace}
\newcommand{\mta}{\textsc{IP}\xspace}
\newcommand{\mov}{\textsc{MR}\xspace}

\newtheorem{theorem}{Theorem}

\newtheorem{lemma}{Lemma}
\newtheorem{mydef}{Definition}

\algdef{SE}[DOWHILE]{Do}{doWhile}{\algorithmicdo}[1]{\algorithmicwhile\ #1}%

\DeclarePairedDelimiter\norm{\lVert}{\rVert}

\title{\pname}
\author{
   Lan N. Nguyen \quad My T. Thai\\
}
\affiliations{

    Department of Computer and Information Science and Engineering \\ 
    University of Florida, Gainesville, Florida 32611 \\
    lan.nguyen@ufl.edu \quad mythai@cise.ufl.edu 
}

\begin{document}

\maketitle

\begin{abstract}
   In this paper, we study a novel problem, \pname (\rfd), as follows: given a ground set $V$; $m$ monotone submodular functions $f_1,...,f_m$; $m$ thresholds $T_1,...,T_m$ and a non-negative integer $r$, \rfd asks for the smallest set $S$ such that for all $i \in [m]$, $\min_{|X| \leq r} f_i(S \setminus X) \geq T_i$. We prove that \rfd is inapproximable within $(1-\epsilon)\ln m$; and no algorithm, taking fewer than exponential number of queries in term of $r$, is able to output a feasible set to \rfd with high certainty. Three bicriteria approximation algorithms with performance guarantees are proposed: one for $r=0$, one for $r=1$, and one for general $r$. We further investigate our algorithms' performance in two applications of \rfd, \pnametap and \pnamemovie. Our algorithms have shown to outperform baseline heuristics in both solution quality and the number of queries in most cases.
\end{abstract}

\section{Introduction}

In a minimum submodular cover, given a ground set $V$, a monotone submodular set function $f: 2^V \rightarrow \mathbb{R}$ and a number $T$, the problem asks for a set $S \subseteq V$ of minimum size such that $f(S) \geq T$. This problem was studied extensively in the literature because of its wide-range applications, e.g. data summarization \cita{mirzasoleiman2015distributed,mirzasoleiman2016fast}, active set selection \cita{norouzi2016efficient}, recommendation systems \cita{guillory2011simultaneous}, information propagation in social networks \cita{kuhnle2017scalable}, and network resilience assessment \cite{nguyen2019network, dinh2014network}. 

However, a single objective function $f$ may not well model several practical applications where achieving multiple goals is required, especially when group fairness is considered. Let us consider the following two representative applications.

\textbf{Information Propagation in Social Network for Multiple Groups}. Social networks are cost-effective tools for information spreading by selecting a set of highly influential people (called seed set) that, through the word-of-mouth effects, the information will be reached to a large number of population \cite{kuhnle2017scalable,nguyen2019influence,zhang2014recent,nguyen2016stop}. For many applications (e.g. broadening participants in STEM), it is important to ensure the diversity and fairness among different ethnics and genders. Therefore, those applications aim to find a minimum seed set such that the information can reach to each group in a fair manner.

\textbf{Items Recommendation for Multiple Users}. Recommendation systems aim to make a good recommendation, e.g. a set of items, which can match users' preferences. In many situations, an item can be served for multiple users, e.g. a family. In this problem, a user's utility level to a set of items is modelled under a monotone submodular function. The objective, therefore, is to find the smallest set of items, from which we can design a recommendation for all users in a way that all reach a certain utility level.





Additionally, these problems require robustness in the solution set, in the sense that the solution satisfies all the constraints even if some elements were removed. Those removal can be from various reasons. For instance, in information propagation, a subset of users may decide not to spread the information \cita{bogunovic2017robust}. Or in recommendation systems, due to the uncertainty of underlying data, information of some items may not be accurate \cita{orlin2018robust}. 

Achieving a reasonable prior distribution on the removed elements may not be practical in many situations. Or even when the distribution is known, it is critical to obtain a robust solution with a high level of certainty, in a way that all goals are still achieved under the worst-case removal. Motivated by that observation, in this work, we study a novel problem, \pname (\rfd), defined as follows.




\begin{mydef} (\rfd)
Given a finite set $V$; $m$ monotone submodular functions $f_1, ... f_m$ where $f_i: 2^V \rightarrow \mathbb{R}^\geq$; $m$ non-negative numbers ${T_1, ..., T_m}$, and a non-negative integer $r$, find a set $S \subseteq V$ of minimum size such that $\forall  ~i \in [m]$, $\min_{|X| \leq r}f_i(S \setminus X) \geq T_i$.
\end{mydef}

\rfd's objective can also be understood as finding $S$ of minimum size such that for all $X \subseteq V$ that $|X| \leq r$ and $i \in [m]$, $f_i(S \setminus X) \geq T_i$. Beside two applications as stated early, \rfd can also be applied in many other applications, such as \textbf{Sensor Placement} \cita{orlin2018robust,ohsaka2015monotone}, which guarantees each measurement (e.g. temperature, humidity) reaches a certain information gain while being robust against sensors' failure; or \textbf{Feature Selection} \cita{qian2017subset,orlin2018robust}, which aims for a smallest set of features that can retain information at a certain level while guaranteeing the set is not dependent on a few features.

To solve \rfd, one direction is to list all constraints in a form of $f_i(. \setminus X) \geq T_i$ $\forall |X| = r$ and $i \in [m]$ and find a smallest set that satisfies all those constraints. However, with large $V$, the amount of set $X$ of size $r$ is $n\choose{r}$, making it impractical to enumerate all possible removed sets. Furthermore, we show that an algorithm, which is able to output a feasible solution to \rfd in general, is very expensive, requiring at least exponential number of queries in term of $r$. Even when $r=0$, there exists no polynomial algorithm that can approximate \rfd within a factor of $(1-\epsilon)\ln m$ unless $P=NP$. Thus solving \rfd remains open and to our knowledge, we are the first one studying the problem.

\textbf{Contribution.} Beside introducing \rfd and investigating the problem's hardness using complexity theory, we propose a bicriteria approximation algorithm, namely \ar, to solve \rfd. \ar's performance guarantee is tight to \rfd's inapproximability and required query complexity. To be specific, \ar is polynomial with fixed $r$ and obtain $S$ of size $O(\ln\frac{\max(m,n)}{\alpha})$ factor to the optimal solution where $\alpha \in (0,1]$. $S$ guarantees that for all $X \subset V$ that $|X| \leq r$ and $i \in [m]$, $f_i(S \setminus X) \geq (1-\alpha) T_i$. In a special case of $r=1$, we propose \ao which can run faster than \ar.

Both \ar and \ao work in a manner that they frequently call an algorithm solving \rfd with $r=0$ as a subroutine. Although \rfd with $r=0$ has been studied in the literature, a new aspect of the problem requires us to propose new solutions to \rfd where $r=0$. 
In particular, we propose Random Greedy (\rg) and re-investigate two existing algorithms, \gr and \thr, whose performance has been analyzed where $f_i$s receive values in $\mathbb{Z}$, in order to adapt them to $\mathbb{R}$ domain. In comparison to \gr and \thr, \rg does not unite submodular functions into a single function; and introduces randomness to reduce queries to $f_i$s. \rg takes much fewer queries than \gr and \thr as shown in our experiments.

Further, we investigate our algorithms' performance on two applications of \rfd: \pnametap and \pnamemovie. The experimental results show our algorithms outperformed some intuitive heuristics methods in both quality of solutions and the number of queries.

\section{Preliminaries} \label{sec:prelim}

\subsection{Related Work} 
To our knowledge, this work provides the first solutions to \rfd for a general $r$. In this part, we pay attention to recent studies on minimum multi-submodular cover (\rfd when $r=0$), and robust submodular optimization.

With minimum submodular cover ($m=1, r=0$), \citet{goyal2013minimizing} showed that the classical greedy algorithm is able to obtain a bi-criteria ratio of $O(\ln\alpha^{-1})$. If we run $m$ instances of greedy, each with a constraint $f_i(.) \geq T_i$, get $m$ output $S_i$ and returns $\cup_{i \in [m]} S_i$, we can get the ratio of $O(m \ln \alpha^{-1})$ for \rfd when $r=0$. In this paper, we aim for algorithms with better ratios.  

\citet{krause2008robust} was the first one proposing a problem of minimum multi-submodular cover; and the problem was then further studied by \citet{mirzasoleiman2016fast,iyer2013submodular}. In general, their solution made a reduction from multiple submodular objectives to a single instance of a submodular cover problem by defining $F(.) = \sum_{i \in [m]} \min(f_i(.), T)$ (all thresholds are the same); and find $S$ of minimum size such that $F(S) = mT$. Two algorithms were proposed, \gr \cita{krause2008robust, iyer2013submodular} and \thr \cita{mirzasoleiman2016fast}. Their performance analysis requires $\{f_i\}_{i \in [m]}$ to receive values in $\mathbb{Z}$ to obtain ratio of $O(\ln \max_{e \in V} F(\{e\}))$. 

However, requiring $\{f_i\}_{i\in[m]}$ to receive values in $\mathbb{Z}$ is not practical in many applications. In our work, we re-investigate \gr and \thr's performance without such requirement. Also, our \rg algorithm differs from such methods in which \rg does not unite objectives into a single function. Furthermore, \rg adds randomness to reduce the query complexity 
while still obtaining an asymptotically equal performance guarantee to that of \gr.

With robust submodular optimization, the concept of finding set that is robust to the removal of $r$ elements was first proposed by \citet{orlin2018robust}. However, their problem is a maximization, namely Robust Submodular Maximization (RSM), defined as follows: Given a ground set $V$, a monotone submodular function $f$, non-negative integers $k$ and $r$, find $S$ s.t $|S| \leq k$ that maximizes $\min_{Z \subseteq A, |Z| \leq r} f(A \setminus Z)$. This problem was later studied further by \citet{bogunovic2017robust,mitrovic2017streaming,staib2019distributionally,anari2019structured}. RSM and \rfd both focus on the worst-case scenario, where the removal of $r$ elements has the greatest impact on the returned solution. Other than that, the two problems are basically different and we are unable to adapt existing algorithms for RSM to solve \rfd with performance guarantees. The key bottleneck preventing us to adapt those algorithms is how to guarantee that a returned solution is robust and satisfied submodular constraints.

\subsection{Definitions \& Complexity}

In this part, we present definitions and theories that would be used frequently in our analysis; and analyze complexity of solving \rfd. Due to page limit, detailed proofs of lemmas and theorems of this part are provided in Appendix.


\begin{mydef} Given an instance of \rfd, including $V, \{f_i\}_{i\in[m]}, \{T_i\}_{i\in[m]}$, a set $A \subseteq V$ is \textbf{$(t,\alpha)$-robust} iff for all $i \in [m]$, $\min_{|X| \leq t} f_i(A \setminus X) \geq (1-\alpha)T_i$.
\end{mydef}
 
Speaking in another way,  \rfd asks us to find a minimum $(r,0)$-robust set. 


Without loss of generality, in our algorithm, we change $f_i(.) := \min(f_i(.) / T_i, 1)$. It is trivial that $f_i$ is still monotone submodular; and \rfd's objective now is to find $S$ that $\min_{|X| \leq r} f_i(S \setminus X) \geq 1$ $\forall~i \in [m]$. 

If there exists a $(r,0)$-robust set, denote $S^*$ as an optimal solution; and $OPT(U,t)$ as a size of the minimum $(t,0)$-robust set that is a subset of $U$ if there is any. So $|S^*| = OPT(V,r)$. We have the following key lemma:

\begin{lemma} \label{lemma:key}
For all $X_1,X_2 \subseteq V$ that $|X_1| = r_1$, $|X_2| = r_2$ and $r_1 + r_2 \leq r$
\begin{align*}
    OPT(V,r) & \geq OPT(V \setminus X_1, r-r_1) \\
    & \geq OPT\big(V \setminus (X_1 \cup X_2), r-r_1-r_2 \big) \\
    & \geq OPT(V,0)
\end{align*}
\end{lemma}

Lemma \ref{lemma:key} is very critical and will be used frequently to obtain performance guarantees of our algorithms. 

Given a \rfd instance and $\alpha \in [0,1]$, we aims to devise algorithms that guarantee:
\begin{itemize}
    \item If there exists $(r,\alpha)$-robust sets in the \rfd instance, the returned solution is $(r,\alpha)$-robust with size at most some factor to $OPT(V,r)$.
    \item Otherwise, the algorithms notify no $(r,0)$-robust set exists.
\end{itemize}

We first study the hardness of devising such an algorithm to solve \rfd. First, we show that: even the sub-task of outputting a $(r,0)$-robust set if there is any, is already very expensive. That is stated in the following theorem.

\begin{theorem} \label{theorem:expo}
    There exists no algorithm, taking fewer than exponential number of queries in term of $r$, is able to verify existence of a $(r,0)$-robust set to \rfd.
\end{theorem}

Theorem \ref{theorem:expo} is proven by taking one instance of \rfd, in which the removal of any subset $X \subseteq V$ of a same size shows a similar behavior on the submodular objectives except for only one unique subset $R$ of size $r$. The thresholds $\{T_i\}_i$ are set so that: if there exists a $(r,0)$-robust set then $V$ is the only $(r,0)$-robust set and $R$ is the only set that would make $V \setminus R$ violate the constraints. Thus any algorithm, taking fewer than $O({|V|\choose{r}})$ queries is unable to verify whether $V$ is $(r,0)$-robust. The full description of the \rfd instance is provided in Appendix.

Furthermore, even there exists $(r,0)$-robust sets, devising approximation algorithms for \rfd is NP-hard. We have the following theorem.

\begin{theorem} \label{theorem:inappox}
There exists no polynomial algorithm that can approximate \rfd, even with $r=0$, within a factor of $(1-\epsilon) \ln m$ given $\epsilon > 0$ unless $P=NP$.
\end{theorem}

\section{Algorithms when $r = 0$} \label{sec:alg_0}

We first study \rfd with $r=0$ since complexity and solution quality of algorithms for \rfd with $r=0$ play critical roles on the performance of \ao and \ar. Although \rfd with $r=0$ has been studied in the literature, these results cannot applied directly. The key barrier is that the initial solution set may not be empty.

In this part, we propose \rg, a randomized algorithm with bicriteria approximation ratio of $O(\ln\frac{m}{\alpha})$. Also, we re-investigate performance guarantees of \gr and \thr, extending from their performance when $f_i$s receive values in $\mathbb{Z}$.

With \rg, checking if there exists feasible solutions with $r=0$ is quite trivial. \rg simply verifies whether $f_i(V) \geq 1-\alpha$ for all $i \in [m]$. If no, the algorithm notifies no feasible set exists and terminates.

      \begin{algorithm}[t]
        \caption{\rg}
    	\label{alg:rand_greedy}
        \begin{flushleft}
        \textbf{Input} $V, S_0, \{f_i\}_{i \in [m]}$ \\
        \end{flushleft}
        \begin{algorithmic}[1]
            \If{There exists $i \in [m]$ s.t. $f_i(V) < 1-\alpha$}
                \State \textbf{Return } no feasible solution
            \EndIf
            \State $t = 0$; $\mathcal{F}_0 = \{f_i\}_{i \in [m]}$
            \While{$\mathcal{F}_t \neq \emptyset$}
                \State $F = $ randomly select $|\mathcal{F}_t|/2$ constraints from $\mathcal{F}_t$
                \State $e_t = \mbox{argmax}_{e \in V \setminus S_t} \sum_{f_i \in F} \Delta_{e} f_i(S_t)$
                \State $S_{t+1} = S_t \cup \{e_t\}$; $\mathcal{F}_{t+1} = \mathcal{F}_t$; $t = t + 1$
                \State Remove all $f_i \in \mathcal{F}_t$ that $f_i(S_t) \geq 1-\alpha$ out of $\mathcal{F}_t$ 
            \EndWhile
        \end{algorithmic}
        \begin{flushleft}
        	\textbf{Return } $S_t$
        \end{flushleft}
      \end{algorithm}

If there exists feasible solutions, \rg works in rounds in order to find a $(0,\alpha)$-robust solution. For each round, a new random process is introduced as follows: the algorithm randomly selects half of functions $f_i$s, each of which is still less than $1-\alpha$; and greedily 
chooses an element that maximizes the sum of marginal gains of the selected functions. This random process helps \rg (1) reduce the number of queries to $f_i$s by half at each round; and (2) establish a recursive relationship of obtained solutions at different rounds, which is critical for \rg to obtain its performance guarantee with high probability (w.h.p).

\rg's pseudocode is presented by Alg. \ref{alg:rand_greedy}. In Alg. \ref{alg:rand_greedy}, $S_t$ represents an obtained solution at round $t$ and $\mathcal{F}_t$ is a set of $f_i$s that $f_j(S_t) \geq 1-\alpha$ $\forall~f_j\not\in \mathcal{F}_t$. Note that \rg starts with $S_0$ as an input; and as can be seen later, \rg is used as a subroutine function in case $r >0$, in which $S_0$ may not be empty. Therefore, analyzing performance of \rg with $S_0 \neq \emptyset$ is necessary and challenging.

To obtain \rg's performance guarantee, we have the following lemma. 
\begin{lemma} \label{lemma:rand_gr}
At round $t$: $\mbox{E}\Big[\sum_{f_i \in \mathcal{F}_{t+1}} \big(1-f_i(S_{t+1}) \big)\Big] \leq \big( 1 - \frac{1}{2~OPT(V,0)} \big) \sum_{f_i \in \mathcal{F}_t} \big(1-f_i(S_{t}) \big)$
\end{lemma}
Lemma \ref{lemma:rand_gr} establishes a recursive relationship between $\mathcal{F}_t$ and $S_t$ at different rounds. This is a key to obtain \rg's approximation ratio. Assuming \rg stops after $L$ rounds, $L = |S_L \setminus S_0|$. By using Markov inequality, we can bound $L$ w.h.p to obtain \rg's performance guarantee as Theorem \ref{theorem:rg_approx}. Full proofs of Lemma \ref{lemma:rand_gr} and Theorem \ref{theorem:rg_approx} are presented in Appendix.
\begin{theorem} \label{theorem:rg_approx}
    Given an instance of \rfd  with input $V,\{f_i\}_{i\in[m]}, S_0$ such that 
    $\sum_{i \in [m]} f_i(S_0) \geq (1-\eta) m$, and $r=0$. If $S$ is an output of \rg then w.h.p $|S \setminus S_0| \leq OPT(V,0) O(\ln\frac{m\eta}{\alpha})$ and each $f_i$ is queried at most $O(|V|OPT(V,0)\ln\frac{m\eta}{\alpha})$ times.
\end{theorem}

We now investigate the performance of \gr and \thr. Their performance guarantees are stated by Theorem \ref{theorem:gr_approx} (\gr) and \ref{theorem:thr_approx} (\thr). Due to page limit, their detailed description and proofs are presented in Appendix. 

\begin{theorem} \label{theorem:gr_approx}
    Given an instance of \rfd  with input $V,\{f_i\}_{i\in[m]}, S_0$ such that $\sum_{i \in [m]} f_i(S_0) \geq (1-\eta) m$, and $r=0$. If \gr terminates with a $(0,\alpha)$-robust solution $S$, 
    then $|S \setminus S_0| \leq OPT(V,0) O(\ln\frac{m\eta}{\alpha})$ and each $f_i$ is queried at most $O(|V|OPT(V,0)\ln\frac{m\eta}{\alpha})$ times.
\end{theorem}

\begin{theorem} \label{theorem:thr_approx}
    Given an instance of \rfd  with input $V,\{f_i\}_{i\in[m]}, S_0$ such that $\sum_{i \in [m]} f_i(S_0) \geq (1-\eta) m$, and $r=0$. If \thr terminates with a $(0,\alpha)$-robust solution $S$, then $|S \setminus S_0| \leq OPT(V,0) O(\frac{1}{1-\gamma}\ln\frac{m\eta}{\alpha})$ where $\gamma \in (0,1)$ is the algorithm's parameter; and each $f_i$ is queried at most $O(\frac{n}{\gamma} \ln \frac{mn}{\alpha})$ times. 
\end{theorem}

With $S_0 = \emptyset$, \rg, \gr and \thr ($\gamma$ is close to $0$) can obtain a ratio of $O(\ln \frac{m}{\alpha})$, which is tight to the inapproximability of \rfd when $r=0$ (Theorem \ref{theorem:inappox}).

\section{Algorithms when $r > 0$} \label{sec:alg_r}

In this section, we propose two algorithms to solve \rfd when $r > 0$: \ao for a special case of $r = 1$ and \ar for general $r$. Both algorithms frequently call an algorithm to \rfd when $r = 0$ as a subroutine, which could be either \rg, \gr or \thr as discussed earlier. In short, we use \az to refer to any of these three.

For simplicity, we ignore the step of notifying if there exists no $(r,\alpha)$ robust set in \ao and \ar's description since it can trivially inferred from the outputs of \az. Without loss of generality, in our analysis, we assume there exists $(r,0)$-robust sets. 

\subsection{Algorithm when $r = 1$ (\ao)}

In general, \ao is an iterative algorithm, which iteratively checks if there exists an element whose removal causes an obtained solution $S$ to violate at least one constraint. If such an element (let's call it $e$) exists, \ao gathers all violated constraints to form a new \rfd instance with $r=0$, $V \setminus \{e\}$ as an input ground set and $S \setminus \{e\}$ as an initial set. This is a key of \ao because by solving that new \rfd instance using \az, \ao guarantees the obtained solution is robust against $e$'s removal; and the algorithm can significantly tighten an upper bound on the number of newly-added elements in order to obtain a tight approximation ratio.

\ao's pseudocode is presented by Alg. \ref{alg:alg_1}. In Alg. \ref{alg:alg_1}, $S_1$ is a $(0,\alpha)$-robust set, found by using \az with the original \rfd's input (line \ref{line:first_out}). $S$, returned by \ao, is $(1,\alpha)$-robust because: 
\begin{itemize}
    \item For any $e \in S_1$ that violates the condition of \textbf{while} loop (line \ref{line:e_1}), \ao guarantees $f_i(S \setminus \{e\}) \geq 1-\alpha$ (output of \az, line \ref{line:e_2}).
    \item For any $e \not\in S_1$, as $S_1 \subseteq S \setminus \{e\}$, we have $f_i(S \setminus \{e\}) \geq f_i(S_1) \geq 1-\alpha$ (output of \az, line \ref{line:first_out}).
\end{itemize}


      \begin{algorithm}[t]
        \caption{Algorithm when $r = 1$ (\ao)}
    	\label{alg:alg_1}
        \begin{flushleft}
        \textbf{Input} $V, \{f_i\}_{i \in [m]}$ \\
        \end{flushleft}
        \begin{algorithmic}[1]
            \State $S = S_1 = $ \az($V, \emptyset, \{f_i\}_{i \in [m]}$) \label{line:a0_first} \label{line:first_out}
            \While{$\exists e \in S_1$ that $\exists~i \in [m]$, $f_i(S \setminus \{e\}) < 1-\alpha$} \label{line:e_1}
                \State $\mathcal{F}^\prime = $ set of all $f_i$ that $f_i(S \setminus \{e\}) < 1-\alpha$
                \State $S^\prime = $ \az($V \setminus \{e\}, S \setminus \{e\}, \mathcal{F}^\prime$) \label{line:e_2}
                \State $S = S \cup S^\prime$
            \EndWhile
        \end{algorithmic}
        \begin{flushleft}
        	\textbf{Return } $S$
        \end{flushleft}
      \end{algorithm}


Denote $E$ as a set of $e \in S_1$ that violate the condition of \textbf{while} loop (line \ref{line:e_1}). For each $e \in E$, denote $S^e$ as $S$ right before $e$ is considered by the \textbf{while} loop of line \ref{line:e_1}. Let $\rho_e = \sum_i \Delta_e f_i(S^e \setminus \{e\}) / m$. To obtain \ao's performance guarantee, we have the following lemma.

\begin{lemma} \label{lemma:observe_alg1}
    $\sum_{e \in E}\rho_e \leq 1$
\end{lemma}

\begin{proof}
Let's sort elements in $S_1 = \{u_1,u_2,...\}$ in the order of being added into $S_1$ by \ao (line \ref{line:first_out}). Let $S_1^e = \{u_1, ..., u_{e-1}\}$. Due to submodularity, $\sum_i \Delta_e f_i(S_1^e) \geq \sum_i \Delta_e f_i(S \setminus \{e\}) = \rho_e m$. Then:
\begin{align*}
    \sum_{e \in E} \rho_e m \leq \sum_{e \in E } \sum_i \Delta_e f_i(S_1^e) \leq \sum_{e \in S_1} \sum_i \Delta_e f_i(S_1^e) \leq m
\end{align*}
which means $\sum_{e \in E} \rho_e \leq 1$ and the proof is completed.
\end{proof}




We then obtain \ao's performance guarantee as stated in Theorem \ref{theorem:alg1_approx}.

\begin{theorem} \label{theorem:alg1_approx}
    Given an instance of \rfd  with input $V,\{f_i\}_{i\in[m]}$ and $r=1$. If $S$ is an output of \ao and $S_1$ is a $(0,\alpha)$-robust set outputted by $\az(V,\emptyset, \{f_i\}_{i\in[m]})$ then $|S| \leq OPT(V,1) O(|S_1| \ln m + 1/\alpha)$. 
\end{theorem}

\begin{proof}
From a ratio of \az and lemma \ref{lemma:key}, we have $|S_1| \leq O(\ln \frac{m}{\alpha}) OPT(V,0) \leq O(\ln \frac{m}{\alpha}) OPT(V,1)$. 

For each $e \in E$ and $S^e$ as defined before, we have:
\begin{align*}
    \sum_i f_i(S^e \setminus \{e\}) = \sum_i f_i(S^e) - \rho_e m \geq m(1 - \alpha - \rho_e)
\end{align*}
The last inequality comes from the fact that $S_1 \subseteq S^e$ and $S_1$ is $(0,\alpha)$-robust. Then, with $S^\prime = $ \az($V \setminus \{e\}, S^e \setminus \{e\}, \mathcal{F}^\prime$) in line \ref{line:e_2}, denote $\delta S^e = S^\prime \setminus S^e $. From the ratio of \az and lemma \ref{lemma:key}, we have:
\begin{align*}
    |\delta S^e| &\leq O(\ln \frac{(\alpha + \rho_e) m}{\alpha}) OPT(V\setminus\{e\},0) \\ & \leq O(\ln \frac{(\alpha + \rho_e) m}{\alpha}) OPT(V,1) 
\end{align*}
Therefore, with $S$ is the returned solution, we have:
\begin{align*}
    |S| & = |S_1| + \sum_{e \in E} | \delta S^e | \\
    & \leq O\Big(\ln \frac{m}{\alpha} + \sum_{e \in E} \ln \frac{(\alpha + \rho_e)m}{\alpha} \Big) OPT(V,1) \\
    & = O\Big(\ln \frac{m}{\alpha} + \ln \prod_{e \in E} \frac{(\alpha + \rho_e)m}{\alpha} \Big) OPT(V,1) \\
    & \leq O\Big(\ln \frac{m}{\alpha} + \ln \big( \sum_{e \in E} \frac{(\alpha+\rho_e)m}{\alpha|E|} \big)^{|E|} \Big) OPT(V,1) \\
    & \leq O\Big(\ln \frac{m}{\alpha} + \ln\big(m (1 + \frac{1}{\alpha |E|}) \big)^{|E|} \Big) OPT(V,1) \\
    & \leq O(|E| \ln m + \frac{1}{\alpha}) OPT(V,1)
\end{align*}
which completes the proof.
\end{proof}

\ao's ratio is \textit{tight} by considering a special instance of \rfd, Robust Set Cover with $r=1$. This tight example is provided in Appendix. 

In term of \textbf{query complexity}, it is trivial that if \ao uses \rg or \gr as \az, each $f_i$ would be queried at most $O(n \max(OPT(V,1) (|S_1| \ln m + 1/\alpha), n))$ times. If \thr is used, each constraint of $\mathcal{F}^\prime$ in line \ref{line:e_2} is queried at most $O(n \ln mn)$ times, thus each $f_i$ is queried at most $O(n |S_1| \ln mn)$ times in total.

\subsection{Algorithm for general $r$ (\ar)}

\ar works in at most $r$ rounds, in which after $t$ rounds, \ar guarantees an obtained solution is $(t,\alpha)$-robust.  Denote $S_t$ as the obtained solution after $t$ rounds. At round $t$, \ar introduces a new \rfd instance with a new set of functions $F_t$. Each function in $F_t$ is defined by a function $f_i$ and a set $X \subset S_t$ that $f_i(S_t \setminus X) < 1-\alpha$ and $|X| = r$. This is a key of \ar because by solving the new \rfd instance to obtain $S_{t+1}$, \rg guarantees $S_{t+1}$ is $(t,\alpha)$-robust. Also the algorithm is able to bound the number of newly-added elements in term of $|S^*|$ by observing that $S^*$ is also a feasible solution to 
the new \rfd instance.

\ar's pseudocode is presented by Alg. \ref{alg:alg_r}. Note that \ar guarantees $S_r$ is $(r,\alpha)$-robust without a need of scanning all the removals of its subsets of size $r$. We prove that by using contradiction as follows: 

Assume $S_r$ is not $(r,\alpha)$-robust, then there exists $X \subseteq V$ and $f_i$ such that $|X| = r$ and $f_i(S_r \setminus X) < 1-\alpha$. Let $X_0 = X \cap S_0$, and $X_t = X \cap (S_t \setminus S_{t-1})$ for $t=1 \rightarrow r$.




If there exists an empty $X_t$, let $X^\prime = \cup_{i=0}^{t-1} X_i$. We have $|X^\prime| \leq r$ and $X^\prime \subseteq S_{t-1}$. Due to the output of \az in line \ref{line:ar_out}, $f_i(S_t \setminus X^\prime) \geq 1-\alpha$. But $S_t \setminus X^\prime \subseteq S_r \setminus X$, so $f_i(S_r \setminus X) \geq 1-\alpha$, which contradicts to our assumption. 

      \begin{algorithm}[t]
        \caption{Algorithm with general $r$ (\ar)}
    	\label{alg:alg_r}
        \begin{flushleft}
        \textbf{Input} $V, \{f_i\}_{i \in [m]}, r$ \\
        \end{flushleft}
        \begin{algorithmic}[1]
            \State $F_0 = \{f_i\}_{i \in [m]}$; $S_0 = $ \az($V, \emptyset, F_0$);
            \For{$t = 1 \rightarrow r$} \label{line:algr_round}
                \State $F_t = \emptyset$
                \For{each set $X \subseteq S_{t-1}$ that $|X| = r, X \not\subseteq S_{t-2}$} \label{line:enumX}
                    \For{each $i \in [m]$ s.t. $f_i(S_{t-1} \setminus X) < 1-\alpha$}
                        \State Define $f_{i,X}(.) = f_i(. \setminus X)$ \label{line:new_con}
                        \State $F_t = F_t \cup \{f_{i,X}\}$
                    \EndFor
                \EndFor
                \State $S_t = $ \az($V, S_{t-1}, F_t$) \label{line:ar_out}
            \EndFor
        \end{algorithmic}
        \begin{flushleft}
        	\textbf{Return } $S_r$
        \end{flushleft}
      \end{algorithm}

Thus, no $X_t$ should be empty, which is impossible since $|X| = r \geq | \cup_{t=0}^r X_t | $, and $X_0, ..., X_r$ are disjoint. Therefore, $S_r$ should be $(r,\alpha)$-robust.

To obtain \ar's performance guarantee, we have the following lemma.
\begin{lemma} \label{lemma:algr_bound}
$|S_t \setminus S_{t-1}| \leq O(\ln \frac{|F_t|}{\alpha}) OPT(V,r)$ for all $t \leq r$
\end{lemma}

\begin{proof}
Considering a new constraint $f_{i,X}$ created in line \ref{line:new_con}, it is trivial that the function $f_{i,X}$ is monotone submodular.


Also, as $S^*$ is $(r,0)$-robust, $f_{i,X}(S^*) = f_i(S^* \setminus X) \geq 1$. That means $S^*$ is feasible for the \rfd instance in line \ref{line:new_con}, with $F_t$ as a set of constraint and $r = 0$. The lemma follows from the ratio of \az.
\end{proof}


Lemma \ref{lemma:algr_bound} is critical to obtain \ar's ratio, stated in the following theorem.


\begin{theorem} \label{theorem:algr_approx}
    Given an instance of \rfd  with input $V,\{f_i\}_{i\in[m]}, r$, if $S$ is an output of \ar, then:
    \begin{align*}
        |S| \leq OPT(V,r) O(r \ln m / \alpha + r^2 \ln n)
    \end{align*}
\end{theorem}

\begin{proof}
Using lemma \ref{lemma:key} and \az's ratio, we have: $|S_0| \leq OPT(V,0) O(\ln m/\alpha) \leq OPT(V,r) O(\ln m/\alpha)$. Therefore, from lemma \ref{lemma:algr_bound}, we have:
\begin{align*}
    |S_r| & = |S_0| + \sum_{t=1}^r |S_t \setminus S_{t-1}| \\
    & \leq O(\ln \frac{m}{\alpha} + \sum_{t=1}^r \ln \frac{|F_t|}{\alpha})~ OPT(V,r)
\end{align*}

Furthermore, $\sum_{t=1}^r |F_t| \leq m \binom{|S_{r-1}|}{r}$ because: (1) No subset $X \in S_{r-1}$ of size $r$ is considered more than one round (line \ref{line:algr_round}) as if $f_i(S_t \setminus X) \leq 1-\alpha$ then $f_i(S_{t+1} \setminus X) \geq 1-\alpha$; and (2) each subset $X$ added to $F_t$ at most $m$ new constraints. 

Therefore, by using AM-GM inequality, we have: $\prod_t |F_t| \leq (\frac{\sum_t |F_t|}{r})^r \leq (\frac{m}{r} \binom{n}{r})^r$.

Thus, $|S_r| \leq O(r \ln \frac{m}{\alpha} + r^2 \ln n)~OPT(V,r)$.
\end{proof}





\textbf{Query Complexity. } The bottleneck of \ar is from the task of finding all subsets $X$ in line $\ref{line:enumX}$. As there is $|S_{r-1}|\choose{r}$ subsets $X$, \ar takes $|S_{r-1}|\choose{r}$ queries for each $f_i$ to only find $X$; and in the worst case, each $f_i$ will generate $|S_{r-1}|\choose{r}$ functions $f_{i,X}$ (line \ref{line:new_con}). Then, if \ar uses \rg or \gr as \az, in worst case, each $f_i$ is queried at most $O(n (r \ln \frac{m}{\alpha} + r^2 \ln n)~OPT(V,r) {|S_{r-1}|\choose{r}})$ times. If \thr is used, at round $t$, each $f_i$ is queried at most $O(\frac{n|F_t|}{\gamma} \ln \frac{n|F_t|}{\alpha})$. Overall, \ar using \thr will query each $f_i$ at most $O(\frac{n}{\gamma} {|S_{r-1}|\choose{r}} (r \ln \frac{m}{\alpha} + r^2 \ln n))$ times. \ar is polynomial with fixed $r$ and  favourable if $OPT(V,r) \ll n$.

\section{Experimental Evaluation} \label{sec:exp}

\begin{figure*}[t]
\begin{tikzpicture}[yscale=0.55, xscale=0.55]
    \begin{groupplot}[group style={group size= 4 by 1}]
        \nextgroupplot[title={$|S|$ (\mta)},xlabel={$T$}, grid style=dashed, grid=both, grid style={line width=.1pt, draw=gray!10}, major grid style={line width=.2pt,draw=gray!50}, every axis plot/.append style={ultra thick, smooth},title style={font=\LARGE}, label style={font=\LARGE}]
                \addplot table [x=t, y=rand_gr, col sep=comma] {data/fb_r_0_T/result.csv}; \label{plot:alg_0_rand_gr}
                \addplot table [x=t, y=gr, col sep=comma] {data/fb_r_0_T/result.csv};
                \label{plot:alg_0_gr}
                \addplot table [x=t, y=thr, col sep=comma] {data/fb_r_0_T/result.csv};
                \label{plot:alg_0_thr}
                \addplot table [x=t, y=separate, col sep=comma] {data/fb_r_0_T/result.csv};
                \label{plot:alg_0_sep}
                \coordinate (top) at (rel axis cs:0,1);
        \nextgroupplot[title={\# queries (\mta)},xlabel={$T$}, grid style=dashed, grid=both, grid style={line width=.1pt, draw=gray!10}, major grid style={line width=.2pt,draw=gray!50}, every axis plot/.append style={ultra thick, smooth},title style={font=\LARGE}, label style={font=\LARGE}]
                \addplot table [x=t, y=rand_gr, col sep=comma] {data/fb_r_0_T/query.csv}; 
                \addplot table [x=t, y=gr, col sep=comma] {data/fb_r_0_T/query.csv};
                \addplot table [x=t, y=thr, col sep=comma] {data/fb_r_0_T/query.csv};
                \addplot table [x=t, y=separate, col sep=comma] {data/fb_r_0_T/query.csv};
        \nextgroupplot[title={$|S|$ (\mov)},xlabel={$T$}, grid style=dashed, grid=both, grid style={line width=.1pt, draw=gray!10}, major grid style={line width=.2pt,draw=gray!50}, every axis plot/.append style={ultra thick, smooth},title style={font=\LARGE}, label style={font=\LARGE}]
                \addplot table [x=t, y=rand_gr, col sep=comma] {data/mv_r_0_T/result.csv}; 
                \addplot table [x=t, y=gr, col sep=comma] {data/mv_r_0_T/result.csv};
                \addplot table [x=t, y=thr, col sep=comma] {data/mv_r_0_T/result.csv};
                \addplot table [x=t, y=separate, col sep=comma] {data/mv_r_0_T/result.csv};
        \nextgroupplot[title={\# queries (\mov)},xlabel={$T$}, grid style=dashed, grid=both, grid style={line width=.1pt, draw=gray!10}, major grid style={line width=.2pt,draw=gray!50}, every axis plot/.append style={ultra thick, smooth},title style={font=\LARGE}, label style={font=\LARGE}]
                \addplot table [x=t, y=rand_gr, col sep=comma] {data/mv_r_0_T/query.csv}; 
                \addplot table [x=t, y=gr, col sep=comma] {data/mv_r_0_T/query.csv};
                \addplot table [x=t, y=thr, col sep=comma] {data/mv_r_0_T/query.csv};
                \addplot table [x=t, y=separate, col sep=comma] {data/mv_r_0_T/query.csv};
                \coordinate (bot) at (rel axis cs:1,0);
    \end{groupplot}
\path (top|-current bounding box.north)--
      coordinate(legendpos)
      (bot|-current bounding box.north);
\matrix[
    matrix of nodes,
    anchor=south,
    draw,
    inner sep=0.2em,
    draw,
    nodes={anchor=center, font=\scriptsize},
  ]at([xshift=-3cm]legendpos)
  {
    \ref{plot:alg_0_rand_gr}& \rg&[5pt]
    \ref{plot:alg_0_gr}& \gr&[5pt]
    \ref{plot:alg_0_thr}& \thr&[5pt]
    \ref{plot:alg_0_sep}& \sep \\};
\end{tikzpicture}
\caption{Performance of algorithms with $r=0$}
 	\label{fig:r_0}
\end{figure*}
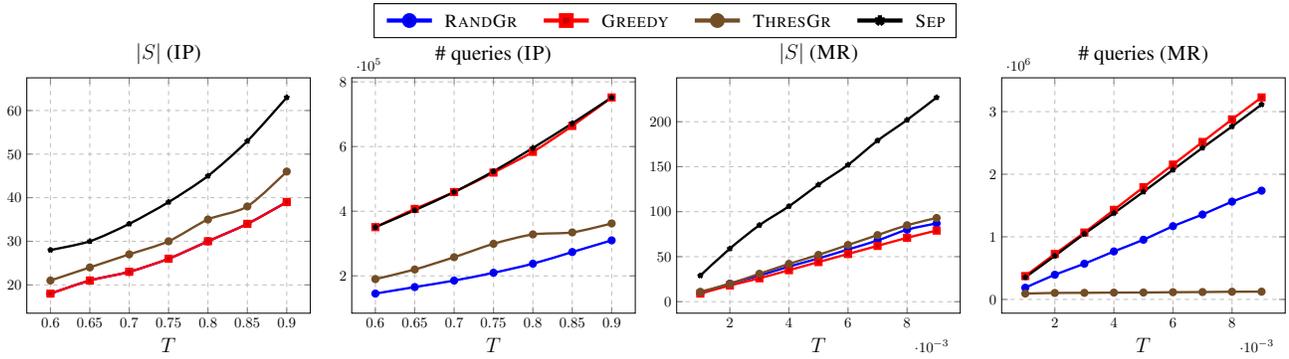

In this section, we compare our algorithms with existing methods and intuitive heuristics on two applications of \rfd, \pnametap (\mta) and \pnamemovie (\mov). The source code is available at \texttt{\url{https://github.com/lannn2410/minrf}}.

\textbf{\pnametap} (\mta) In this problem, a social network is modeled as a directed graph $G=(V,E)$ where $V$ is a set of social users. Each edge $(u,v)$ is associated with a weight $w_{u,v}$, representing the strength of influence from user $u$ to $v$. 

To model the information propagation process, we use Linear Threshold (LT) Model \cite{kempe2003maximizing, nguyen2020streaming}. In general, the process is as follows: Each $v \in V$ has a threshold $\theta_v$ chosen uniformly at random in $[0,1]$ and the information start from a seed set $S \subseteq V$. At first all users in $S$ become active. Next, information cascades in discrete steps and in each step, a user $v$ becomes active if $\sum_{\mbox{active } u} w_{u,v} \geq \theta_v$. The process stops when no more user can become active.


Given a collection $\mathcal{U}$ of subsets of $V$, i.e $\mathcal{U} = \{C_1,...,C_m\}$ where $C_i \subseteq V$. Each $C_i$ represents a group that we need to influence. Denote $I_i(S)$ as the expected number of active users in $C_i$ by a seed set $S$. Given a number $T \in [0,1]$, \mta aims to find the smallest $S$ such that for all $C_i \in \mathcal{U}$, $\min_{|X| \leq r} I_i(S \setminus X) \geq T |C_i|$.


We use Facebook dataset from SNAP database \cita{snapnets}, an undirected graph with 4,039 nodes and 88,234 edges. Since it is undirected, we treat each edge as two directed edges. The weight $w_{u,v}$ is set to be $1/d_v$ where $d_v$ is in-degree of $v$. $\mathcal{U}$ is a collection of groups to which users are classified based on their gender or race. 
Due to lack of data information, a user's race and gender are randomly assigned. $I_i(S)$ is estimated over 100 graph samples.   


\textbf{Movie Recommendation for Multiple Users} (\mov) In this problem, given a set $M$ of movies, a set $U$ of users, each user $u$ has a list $L_u$ of his/her favourite movies. Given $S \subseteq U$, a utility score of $u$ to $S$ is defined as $f_u(S) = \sum_{i \in L_u; j \in S \setminus L_u} s_{i,j}$ \cita{mirzasoleiman2016fast} where $s_{i,j} \in [0,1]$ which measures the similarity between movie $i$ and $j$. Given a number $T$, the objective is to find the smallest set of movies to recommend to all users in a way such that every user's utility level is at least $T$ under any $r$ ``inaccurate-data" movies removal, i.e. $\min_{|X| \leq r} f_u(S \setminus X) \geq T$ for all $u \in U$.



We use Movie Lens dataset from \citet{moviedata} database, which includes information of 10,381 movies; and their 20,000,264 ratings (ranging in $[0,5]$) from 138,493 users. We randomly pick 4 users for a set $U$, $L_u$ contains movies that $u$ rated at least 4. Each movie $i$ is associated by a 1,129-dimension vector $v_i$, where each entry (ranging in $[0,1]$) represents the relevant score between the movie and a keyword. The relevant scores are available in the dataset. We use cosine similarity score $\frac{v_i \cdot v_j}{\norm{v_i}\norm{v_j}}$ to present $s_{i,j}$. For each user $u$, $f_u(S)$ is normalized to be in range $[0,1]$.

\textbf{Compared Algorithms} With $r = 0$, we compare \rg, \gr and \thr ($\gamma = 0.2$) with \sep algorithm: which considers each constraint separately, runs greedy to find a set $S_i$ that $f_i(S_i) \geq 1-\alpha$ and return $\cup_{i\in[m]} S_i$. \sep obtains a ratio of $O(m \ln \frac{1}{\alpha})$. 

With $r > 0$, we compare \ar's performance in combination with each \az, including \rg, \gr, \thr, \sep. Each combination of \ar to a \az algorithm is denoted, in short, \ar-name of the \az algorithm, e.g. \ar-\rg.

We also compare \ar with \djt, a heuristic we propose to evaluate. \djt finds $r+1$ disjoint sets $S_1, ... ,S_{r+1}$ such that $f_i(S_j) \geq 1-\alpha$ for all $i \in [m]$ and $j \in [r+1]$; and returns $S = \cup_{j\in[r+1]} S_j$. If \djt successfully finds all $\{S_j\}_{j\in [r+1]}$, then $S$ is feasible to \rfd without the need for checking all subsets of size $r$. This is because for any set $X$ of size $r$, there should exist $S_j$ that $S_j \cap X = \emptyset$. Thus, $S_j \subset S \setminus X$, which means $f_i(S \setminus X) \geq f_i(S_j) \geq 1-\alpha$ for all $i\in [m]$. However, there are two problems with \djt: (1) If \djt cannot find all $\{S_j\}_{j\in[r+1]}$, the algorithm does not guarantee there exists no feasible solution to \rfd; and (2) \djt does not obtain any approximation ratio.  

For $r=1$, we also evaluate \ao performance in combination with each \az algorithm, including \rg, \gr, \thr. 


\textbf{Other}. We set $\alpha = 0.1$. Results are averaged over 10 repetitions. 

\begin{figure*}[t]
\begin{tikzpicture}[yscale=0.55, xscale=0.55]
    \begin{groupplot}[group style={group size= 4 by 1}]
        \nextgroupplot[title={$|S|$ (\mta)},xlabel={$T$}, grid style=dashed, grid=both, grid style={line width=.1pt, draw=gray!10}, major grid style={line width=.2pt,draw=gray!50}, every axis plot/.append style={ultra thick, smooth},title style={font=\LARGE}, label style={font=\LARGE}]
                \addplot table [x=t, y=alg1_rand_gr, col sep=comma] {data/fb_r_1_T/result.csv}; \label{plot:r_1_alg_1_rand_gr}
                \addplot table [x=t, y=alg1_gr, col sep=comma] {data/fb_r_1_T/result.csv};
                \label{plot:r_1_alg_1_gr}
                \addplot table [x=t, y=alg1_thr, col sep=comma] {data/fb_r_1_T/result.csv};
                \label{plot:r_1_alg_1_thr}
                \addplot table [x=t, y=algR_rand_gr, col sep=comma] {data/fb_r_1_T/result.csv};
                \label{plot:r_1_alg_r_rand_gr}
                \addplot[orange,mark=o] table [x=t, y=algR_gr, col sep=comma] {data/fb_r_1_T/result.csv};
                \label{plot:r_1_alg_r_gr}
                \addplot[cyan,mark=o] table [x=t, y=algR_thr, col sep=comma] {data/fb_r_1_T/result.csv};
                \label{plot:r_1_alg_r_thr}
                \addplot[gray,mark=x] table [x=t, y=separate, col sep=comma] {data/fb_r_1_T/result.csv};
                \label{plot:r_1_separate}
                \addplot[green,mark=o] table [x=t, y=disjoint, col sep=comma] {data/fb_r_1_T/result.csv};
                \label{plot:r_1_disjoint}
                \coordinate (top) at (rel axis cs:0,1);
        \nextgroupplot[title={\# queries (\mta)},xlabel={$T$}, grid style=dashed, grid=both, grid style={line width=.1pt, draw=gray!10}, major grid style={line width=.2pt,draw=gray!50}, every axis plot/.append style={ultra thick, smooth},title style={font=\LARGE}, label style={font=\LARGE}]
                \addplot table [x=t, y=alg1_rand_gr, col sep=comma] {data/fb_r_1_T/query.csv};
                \addplot table [x=t, y=alg1_gr, col sep=comma] {data/fb_r_1_T/query.csv};
                \addplot table [x=t, y=alg1_thr, col sep=comma] {data/fb_r_1_T/query.csv};
                \addplot table [x=t, y=algR_rand_gr, col sep=comma] {data/fb_r_1_T/query.csv};
                \addplot[orange,mark=o] table [x=t, y=algR_gr, col sep=comma] {data/fb_r_1_T/query.csv};
                \addplot[cyan,mark=o] table [x=t, y=algR_thr, col sep=comma] {data/fb_r_1_T/query.csv};
                \addplot[gray,mark=x] table [x=t, y=separate, col sep=comma] {data/fb_r_1_T/query.csv};
                \addplot[green,mark=o] table [x=t, y=disjoint, col sep=comma] {data/fb_r_1_T/query.csv};
        \nextgroupplot[title={$|S|$ (\mov)},xlabel={$T$}, grid style=dashed, grid=both, grid style={line width=.1pt, draw=gray!10}, major grid style={line width=.2pt,draw=gray!50}, every axis plot/.append style={ultra thick, smooth},title style={font=\LARGE}, label style={font=\LARGE}]
                \addplot table [x=t, y=alg1_rand_gr, col sep=comma] {data/mv_r_1_T/result.csv};
                \addplot table [x=t, y=alg1_gr, col sep=comma] {data/mv_r_1_T/result.csv};
                \addplot table [x=t, y=alg1_thr, col sep=comma] {data/mv_r_1_T/result.csv};
                \addplot table [x=t, y=algR_rand_gr, col sep=comma] {data/mv_r_1_T/result.csv};
                \addplot[orange,mark=o] table [x=t, y=algR_gr, col sep=comma] {data/mv_r_1_T/result.csv};
                \addplot[cyan,mark=o] table [x=t, y=algR_thr, col sep=comma] {data/mv_r_1_T/result.csv};
                \addplot[gray,mark=x] table [x=t, y=separate, col sep=comma] {data/mv_r_1_T/result.csv};
                \addplot[green,mark=o] table [x=t, y=disjoint, col sep=comma] {data/mv_r_1_T/result.csv};
        \nextgroupplot[title={\# queries (\mov)},xlabel={$T$},title style={font=\LARGE}, label style={font=\LARGE}, grid style=dashed, grid=both, grid style={line width=.1pt, draw=gray!10}, major grid style={line width=.2pt,draw=gray!50}, every axis plot/.append style={ultra thick, smooth}]
                \addplot table [x=t, y=alg1_rand_gr, col sep=comma] {data/mv_r_1_T/query.csv};
                \addplot table [x=t, y=alg1_gr, col sep=comma] {data/mv_r_1_T/query.csv};
                \addplot table [x=t, y=alg1_thr, col sep=comma] {data/mv_r_1_T/query.csv};
                \addplot table [x=t, y=algR_rand_gr, col sep=comma] {data/mv_r_1_T/query.csv};
                \addplot[orange,mark=o] table [x=t, y=algR_gr, col sep=comma] {data/mv_r_1_T/query.csv};
                \addplot[cyan,mark=o] table [x=t, y=algR_thr, col sep=comma] {data/mv_r_1_T/query.csv};
                \addplot[gray,mark=x] table [x=t, y=separate, col sep=comma] {data/mv_r_1_T/query.csv};
                \addplot[green,mark=o] table [x=t, y=disjoint, col sep=comma] {data/mv_r_1_T/query.csv};
                \coordinate (bot) at (rel axis cs:1,0);
    \end{groupplot}
\path (top|-current bounding box.north)--
      coordinate(legendpos)
      (bot|-current bounding box.north);
\matrix[
    matrix of nodes,
    anchor=south,
    draw,
    inner sep=0.2em,
    draw,
    nodes={anchor=center, font=\scriptsize},
  ]at([xshift=-3cm]legendpos)
  {
    \ref{plot:r_1_alg_1_rand_gr}& \ao \rg&[5pt]
    \ref{plot:r_1_alg_1_gr}& \ao \gr&[5pt]
    \ref{plot:r_1_alg_1_thr}& \ao \thr&[5pt]
    \ref{plot:r_1_separate}& \ar \sep \\
    \ref{plot:r_1_alg_r_rand_gr}& \ar \rg &[5pt]
    \ref{plot:r_1_alg_r_gr}& \ar \gr&[5pt]
    \ref{plot:r_1_alg_r_thr}& \ar \thr&[5pt]
    \ref{plot:r_1_disjoint}& \djt \\};
\end{tikzpicture}
\caption{Performance of algorithms with $r=1$}
 	\label{fig:r_1}
\end{figure*}
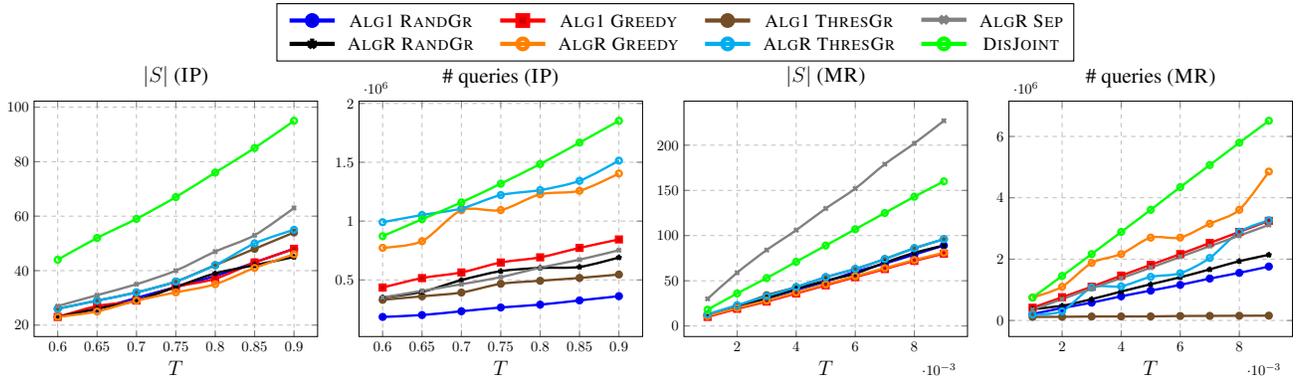

\begin{figure*}[t]
\centering
\begin{tikzpicture}[yscale=0.55, xscale=0.55]
    \begin{groupplot}[group style={group size= 4 by 1}]
        \nextgroupplot[title={$|S|$ (\mta)},title style={font=\LARGE}, label style={font=\LARGE},xlabel={$r$}, grid style=dashed, grid=both, grid style={line width=.1pt, draw=gray!10}, major grid style={line width=.2pt,draw=gray!50}, every axis plot/.append style={ultra thick, smooth},xtick distance=1]
                \addplot table [x=r, y=rand_gr, col sep=comma] {data/fb_R/result.csv}; \label{plot:alg_r_rand_gr}
                \addplot table [x=r, y=greedy, col sep=comma] {data/fb_R/result.csv};
                \label{plot:alg_r_gr}
                \addplot table [x=r, y=threshold, col sep=comma] {data/fb_R/result.csv};
                \label{plot:alg_r_th}
                \addplot table [x=r, y=separate, col sep=comma] {data/fb_R/result.csv};
                \label{plot:alg_r_sep}
                \addplot[green,mark=o] table [x=r, y=disjoint, col sep=comma] {data/fb_R/result.csv};
                \label{plot:alg_r_disjoint}
                \coordinate (top) at (rel axis cs:0,1);
        \nextgroupplot[title={\# queries (\mta)},xlabel={$r$},title style={font=\LARGE}, label style={font=\LARGE}, grid style=dashed, grid=both, grid style={line width=.1pt, draw=gray!10}, major grid style={line width=.2pt,draw=gray!50}, every axis plot/.append style={ultra thick, smooth},xtick distance=1]
                \addplot table [x=r, y=rand_gr, col sep=comma] {data/fb_R/query.csv}; 
                \addplot table [x=r, y=greedy, col sep=comma] {data/fb_R/query.csv};
                \addplot table [x=r, y=threshold, col sep=comma] {data/fb_R/query.csv};
                \addplot table [x=r, y=separate, col sep=comma] {data/fb_R/query.csv};
                \addplot[green,mark=o] table [x=r, y=disjoint, col sep=comma] {data/fb_R/query.csv};
        \nextgroupplot[title={$|S|$ (\mov)},title style={font=\LARGE}, label style={font=\LARGE},xlabel={$r$}, grid style=dashed, grid=both, grid style={line width=.1pt, draw=gray!10}, major grid style={line width=.2pt,draw=gray!50}, every axis plot/.append style={ultra thick, smooth}]
                \addplot table [x=r, y=rand_gr, col sep=comma] {data/mv_R/result.csv}; 
                \addplot table [x=r, y=greedy, col sep=comma] {data/mv_R/result.csv};
                \addplot table [x=r, y=thr, col sep=comma] {data/mv_R/result.csv};
                \addplot table [x=r, y=separate, col sep=comma] {data/mv_R/result.csv};
                \addplot[green,mark=o] table [x=r, y=disjoint, col sep=comma] {data/mv_R/result.csv};
        \nextgroupplot[title={\# queries (\mov)},title style={font=\LARGE}, label style={font=\LARGE},xlabel={$r$}, grid style=dashed, grid=both, grid style={line width=.1pt, draw=gray!10}, major grid style={line width=.2pt,draw=gray!50}, every axis plot/.append style={ultra thick, smooth}, ]
                \addplot table [x=r, y=rand_gr, col sep=comma] {data/mv_R/query.csv}; 
                \addplot table [x=r, y=greedy, col sep=comma] {data/mv_R/query.csv};
                \addplot table [x=r, y=thr, col sep=comma] {data/mv_R/query.csv};
                \addplot table [x=r, y=separate, col sep=comma] {data/mv_R/query.csv};
                \addplot[green,mark=o] table [x=r, y=disjoint, col sep=comma] {data/mv_R/query.csv};
                \coordinate (bot) at (rel axis cs:1,0);
    \end{groupplot}
\path (top|-current bounding box.north)--
      coordinate(legendpos)
      (bot|-current bounding box.north);
\matrix[
    matrix of nodes,
    anchor=south,
    draw,
    inner sep=0.2em,
    draw,
    nodes={anchor=center, font=\scriptsize},
  ]at([xshift=-3cm]legendpos)
  {
    \ref{plot:alg_r_rand_gr}&  \ar \rg&[5pt]
    \ref{plot:alg_r_gr}&  \ar \gr&[5pt]
    \ref{plot:alg_r_th}&  \ar \thr&[5pt]
    \ref{plot:alg_r_sep}&  \ar \sep&[5pt]
    \ref{plot:alg_r_disjoint}& \djt \\};
\end{tikzpicture}
\caption{Performance of algorithms with various $r$}
 	\label{fig:r}
\end{figure*}
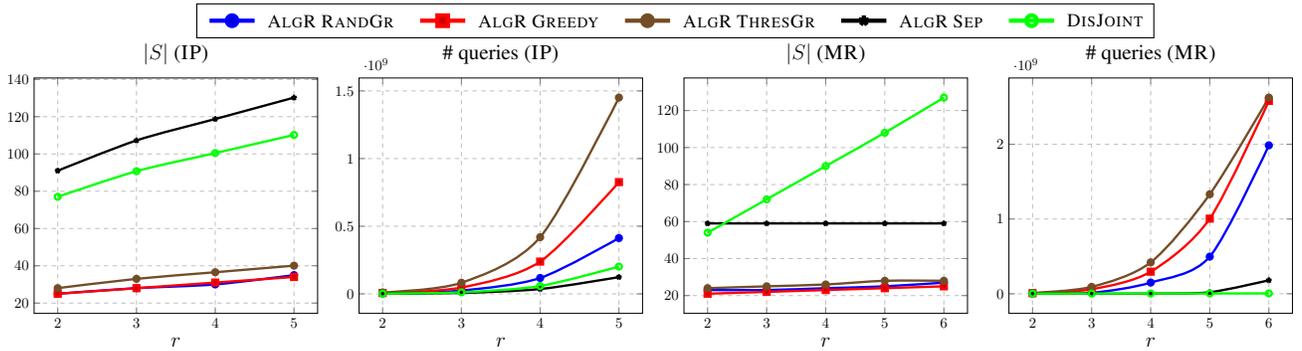

\subsection{Experimental Results}

Fig. \ref{fig:r_0} shows the performances of different \az algorithms in comparison with \sep. We can see that \az algorithms totally outperformed \sep in solution quality by a huge margin. \rg returned solutions approximately close to \gr, which is the best one in term of solution quality. However, in term of query efficiency, \rg took much fewer queries than \gr and; and was the fastest algorithm in the \mta problem. This confirms the efficiency of \rg by introducing randomness and discarding satisfied constraints after each iteration. 

Fig. \ref{fig:r_1} shows algorithms' performance on the \mta and \mov problems when $r = 1$. The two proposed heuristics, \ao-\sep and \djt, showed the worst performance in solution quality. \djt's undesirable performance came from the fact that a 
union of disjoint subsets, each is able to satisfy all constraints, is not a necessary condition to guarantee robustness. Also, by finding disjoint subsets, \djt needed more queries than any other algorithms.

In combination with the same \az algorithm, \ao and \ar had almost similar returned solution but \ao totally outperformed \ar in term of number of queries. That can be explained by the fact that whenever \ao finds an element $e$ whose removal violates at least one constraint, \ao will add elements to compensate for $e$'s removal. That guarantees not only $S$ is robust to $e$'s removal but also the newly-added elements may help $S$ being robust against some other elements' removal as well. On the other hand, \ar gathers all elements, each element's removal violates at least one constraint, to form a new \rfd instance with a much larger set of submodular functions than \ao. That helps \ao obtain better number of queries than \ar.

Fig. \ref{fig:r} shows algorithms' performance with larger $r$. We observed that \ar-\sep and \djt were outperformed by other algorithms by a huge margin in solution quality; but took much fewer number of queries than the others. That is because with larger $r$, the number of subsets of size $r$ is increased by an exponent rate in term of $r$, which increases significantly the number of queries of \ar for scanning subsets of size $r$ of $S_{r-1}$. \sep was less suffered than our \az algorithms because \sep returned much larger solutions, which can reach robustness at $S_t$ where $t \ll r$. On the other hand, \djt had the small number of queries because \djt does not need to scan all removals of its subsets of size $r$ to check feasibility of the returned solution.


Fig. \ref{fig:r} also shows that: \ar-\rg performed the best in both solution quality and the number of queries in comparison with \ar-\gr and \ar-\thr. Although \thr was the most efficient \az algorithm when $r=0$ (standalone) or $r=1$ (combining with \ao or \ar), \ar-\thr's performances were undesirable with large $r$. This is because \thr tends to return larger solution than \rg and \gr. Therefore, \ar-\thr requires more queries to scan over all subset of size $r$ of $S_{r-1}$ than \ar-\rg and \ar-\gr.


  \section{Conclusion}  \label{sec:conclusion}

Motivated by real-world applications, in this work, we studied a problem of minimum robust set subject to multiple submodular constraints, namely \rfd. We investigate \rfd's hardness using complexity theories; and proposed multiple approximation algorithms to solve \rfd. Our algorithms are proven to return tight performance guarantees to \rfd's inapproximability and required query complexity. Finally, we empirically demonstrated that our algorithms outperform several intuitive methods in terms of the solution quality and number of queries.

\section*{Acknowledgements}\label{sc:acknowledgements}
This work was supported in part by the National Science Foundation (NSF) grants IIS-1908594, IIS-1939725, and the University of Florida Informatics Institute Fellowship Program. We would like to thank the anonymous reviewers for their helpful feedback.

\bibliography{reference}


\clearpage
\onecolumn
\appendix





\section{Hardness and complexity requirement of \rfd} \label{apd:prelim}
\subsection{Proof of Lemma \ref{lemma:key}}
\begin{proof}
We first focus on the second inequality since the first inequality can be trivially inferred by the second one. Let $l = |OPT(V \setminus X_1, r-r_1) \cap X_2| \leq r_2$. Since $OPT(V \setminus X_1, r-r_1)$ is robust to a removal of $r-r_1$ elements, $OPT(V \setminus X_1, r-r_1) \setminus X_2$ is robust to a removal of $r-r_1-r_2$ elements and contains no elements from $X_2$. Therefore,  $OPT(V \setminus X_1, r-r_1) \setminus X_2$ is a feasible solution given input $V \setminus X_1 \setminus X_2 = V \setminus (X_1 \cup X_2)$ and $r-r_1-r_2$. So $OPT(V \setminus X_1, r-r_1) \geq OPT(V \setminus (X_1 \cup X_2), r-r_1-r_2)$.

The last inequality comes from observation that $f_i(OPT(V \setminus (X_1 \cup X_2), r-r_1-r_2)) \geq T_i$ for all $i \in [m]$, thus it is feasible to input $V$ and $r=0$.
\end{proof}

\subsection{Proof of Theorem \ref{theorem:expo}}
\begin{proof}
The main idea of this proof is to find a submodular function $f$ and a threshold $T$ such that the removal of any subset $X \subseteq S$ of the same size on $V$ shows a similar behavior on $f$ except for only one unique subset $R$ of size $r$. $R$ is the only set that $f(V \setminus R) < T$. Thus any algorithm, taking fewer than $O({|V|\choose{r}})$ queries is unable (or only with tiny probability) to  verify whether $V$ is $(r,0)$-robust or not. 

The instance is as follows: Given the ground set $V$ and $r$, we randomly choose a subset $R \subset V$ that $|R| = r$. The submodular function $f$ is defined as follows:
\begin{itemize}
    \item For any $Z \subseteq V$ that $|Z| < |V| - r$, $f(Z) = 3|Z|$. 
    \item $f(V \setminus R) = 3 |V \setminus R| - 1$
    \item For any $Z \subseteq V$ that $|Z| \geq |V| - r$ and $Z \neq V \setminus R$, $f(Z) = 3(|V| - r) + (|Z| - |V| + r)$
\end{itemize}

It is trivial that $f$ is monotone. We now prove that $f$ is submodular. Given $A \subset B \subseteq V$ and $e \not\in B$, we have:
\begin{itemize}
    \item If $|A| < |V| - r$ then:
    \begin{itemize}
        \item If $A \cup \{e\} = V \setminus R$, $\Delta_e f(A) = 2$ while $\Delta_e f(B) = 1$
        \item Otherwise $\Delta_e f(A) = 3$ while
        \begin{itemize}
            \item $\Delta_e f(B) = 3$ if $| B \cup \{e\} | \leq |V| - r$ and $B \cup \{e\} \neq V\setminus R$
            \item $\Delta_e f(B) = 2$ if $B \cup \{e\} = V\setminus R $ or $B = V \setminus R$
            \item Otherwise, $\Delta_e f(B) = 1$
        \end{itemize}
    \end{itemize}
    \item If $|A| \geq |V| - r$, then
    \begin{itemize}
        \item If $A = V \setminus R$, $\Delta_e f(A) = 2$ while $\Delta_e f(B) = 1$
        \item Otherwise $\Delta_e f(A) = \Delta_e f(B) = 1$
    \end{itemize}
\end{itemize}
So, in any cases, $\Delta_e f(A) \geq \Delta_e f(B)$. Thus, $f$ is submodular.

Let $T = 3(|V| - r)$. Then $R$ is an only set of size $r$ that satisfies $f(V \setminus R) < T$. Thus, any algorithm making sub exponentially many queries will be unable (except with tiny probability) to find $R$.
\end{proof}

\subsection{Proof of Theorem \ref{theorem:inappox}}
\begin{proof}
We reduce SET COVER to \rfd with $r=0$. 

The SET COVER problem is: Given a finite set $V^\prime$ and a collection $\mathcal{C}$ of subset $S_1,...S_l$ ($S_i \subseteq V^\prime$), find $\mathcal{S} \subseteq \mathcal{C}$ of minimum size such that $\cup_{S_i \in \mathcal{S}} S_i = V$. SET COVER can be formulated by the following Integer Programming.

\begin{align} \label{equ:ip_set_cover}
    \mbox{Minimize } \sum_{i = 1}^l x_i \quad \quad \mbox{s.t. } \sum_{i : e \in S_i} x_i \geq 1 \quad \forall~ e\in V^\prime; \mbox{ and } x_i \in \{0,1\} \quad \forall~ i=1,...,l 
\end{align}

To reduce it to an instance of \rfd, we define $V = \mathcal{C}$. For each $e \in V^\prime$, define $f_e(\mathcal{S})$ as number of sets in $\mathcal{S}$ that contains $e$. $f_e$ is not only submodular, but modular. $T_e = 1$ for all $e \in V^\prime$. Then solving the above Integer Programming is equivalent to finding minimum $\mathcal{S}$ that $f_e(\mathcal{S}) \geq T_e$ for all $e \in V^\prime$. In this \rfd instance, $m = |V^\prime|$.

If there exists a $(1-\epsilon) \ln m$ approximation algorithm $\mathcal{A}$ for \rfd, which means we can use $\mathcal{A}$ to approximate SET COVER within $(1-\epsilon) \ln |V^\prime|$ ratio. That contradicts with \citet{dinur2014analytical} that SET COVER is inapproximable within ratio $(1-\epsilon) \ln |V^\prime|$ unless $P=NP$.
\end{proof}

\section{Omitted proofs of \rg} \label{apd:rg}


\subsection{Proof of Lemma \ref{lemma:rand_gr}}
\begin{proof}
Considering at round $t$ and $F$ is a set of randomly selected constraints, we have:
\begin{align}
    \sum_{f_i \in F} (1 - f_i(S_t)) & \leq \sum_{f_i \in F} (f_i(S^* \cup S_t) - f_i(S_t)) \leq \sum_{f_i \in F} \sum_{e \in S^* \setminus S_t} \Delta_e f_i(S_t) \\
    & \leq \sum_{e \in S^* \setminus S_t} \sum_{f_i \in F}  \Delta_e f_i(S_t) \leq OPT(V,0)  \sum_{f_i \in F}  \Delta_{e_t} f_i(S_t) \label{equ:rgr_1}
\end{align}
Let $\mathcal{H}$ is a set of all combination of size $|\mathcal{F}_t| / 2$ of $\mathcal{F}_t$. For each $F \in \mathcal{H}$, denote $e_{t,F}$ as $e_t$ if $F$ is selected. Then,
\begin{align}
    \mbox{E}_{F \sim \mathcal{H}}[\sum_{f_i \in F} (1 - f_i(S_t))] & =  \sum_{F \in \mathcal{H}} \frac{1}{|\mathcal{H}|} \sum_{f_i \in F} (1 - f_i(S_t)) \\
    & = \frac{1}{|\mathcal{H}|} \sum_{f_i \in \mathcal{F}_t} \sum_{F \in H: f_i \in F} (1 - f_i(S_t)) \\
    & = \frac{1}{2} \sum_{f_i \in \mathcal{F}_t} (1 - f_i(S_t)) \label{equ:rgr_2}
\end{align}

On the other hand,
\begin{align}
    \mbox{E}_{F \sim \mathcal{H}}\Big[OPT(V,0) \sum_{f_i \in F} \Delta_{e_{t,F}} f_i(S_t) \Big] & =  \sum_{F \in \mathcal{H}} \frac{1}{|\mathcal{H}|} OPT(V,0) \sum_{f_i \in F} \Delta_{e_{t,F}} f_i(S_t) \\
    & = OPT(V,0) \sum_{f_i \in \mathcal{F}_t}  \frac{1}{|\mathcal{H}|} \sum_{F \in \mathcal{H} : f_i \in F} \Delta_{e_{t,F}} f_i(S_t) \\
    & \leq OPT(V,0) \sum_{f_i \in \mathcal{F}_t}  \frac{1}{|\mathcal{H}|} \sum_{F \in \mathcal{H}} \Delta_{e_{t,F}} f_i(S_t) \\
    & = OPT(V,0) \sum_{f_i \in \mathcal{F}_t} \mbox{E}\Big[\Delta_{e_t} f_i(S_t) \Big] \label{equ:rgr_3}
\end{align}

Combining (\ref{equ:rgr_1}), (\ref{equ:rgr_2}), (\ref{equ:rgr_3}) and the fact that $\mathcal{F}_{t+1} \subseteq \mathcal{F}_t$ and $f_i(S) \leq 1$, we have:
\begin{align*}
    \mbox{E}\Big[\sum_{f_i \in \mathcal{F}_{t+1}} (1-f_i(S_{t+1}))\Big] \leq \sum_{f_i \in \mathcal{F}_t} (1-\mbox{E}[f_i(S_{t+1})]) \leq \bigg( 1 - \frac{1}{2~OPT(V,0)} \bigg) \sum_{f_i \in \mathcal{F}_t} (1-f_i(S_{t}))
\end{align*}
which completes the proof.
\end{proof}

\subsection{Proof of Theorem \ref{theorem:rg_approx}}
\begin{proof}
    From Lemma. \ref{lemma:rand_gr}, after adding $L$ elements, \rg guarantees:
\begin{align*}
    \mbox{E}\Big[\sum_{f_i \in \mathcal{F}_L} (1-f_i(S_L))\Big] \leq \bigg( 1 - \frac{1}{2~OPT(V,0)} \bigg)^L \sum_{f_i \in \mathcal{F}_0} (1-f_i(S_0)) \leq e^{-\frac{L}{2~OPT(V,0)}} m\eta
\end{align*}

Let's consider the probability the algorithm cannot terminate after adding $L$ elements. That probability is equal to the probability that there exists $f_i \in \mathcal{F}_L$ that $f_i(S_L) < 1-\alpha$. We have:
\begin{align*}
    \mbox{Pr}[\exists~f_i\in \mathcal{F}_{L}~\mbox{that}~f_i(S_L) < 1-\alpha] & \leq \mbox{Pr}[\sum_{f_i \in \mathcal{F}_{L}} (1-f_i(S_L)) > \alpha] \\
    & \leq^{(*)} \frac{\mbox{E}\Big[\sum_{f_i \in \mathcal{F}_{L}} (1-f_i(S_L))\Big]}{\alpha} \\
    & \leq e^{-\frac{L}{2~OPT(V,0)}} \frac{m\eta}{\alpha}
\end{align*}
where the inequality $(*)$ is from Markov inequality. 


Therefore, with high probability $1-o(1)$, the algorithm terminates after adding $L = OPT(V,0) \Theta(\ln\frac{m\eta}{\alpha})$ elements. Which also means: With $S_0 = \emptyset$, \rg obtains ratio of $O(\ln \frac{m}{\alpha})$ w.h.p and each $f_i$ is queries by at most $O(|V|OPT(V,0)\ln\frac{m}{\alpha})$. 
\end{proof}

\section{\gr and \thr} \label{apd:gr}

In general, \gr and \thr contain the following steps:
\begin{enumerate}
    \item Set $F(.) = \sum_{i \in [m]} f_i(.)$; $t=0$
    \item While there exists $f_i$ that $f_i(S_t) < 1-\alpha$
    \begin{enumerate}
        \item Find $e_t \in V \setminus S_t$ that $\Delta_{e_t} F(S_t) \geq \delta \times \max_{e \in V \setminus S_t} \Delta_{e} F(S_t)$
        \item $S_{t+1} = S_t \cup \{e_t\}$; $t = t+1$
    \end{enumerate}
    \item Return $S_t$
\end{enumerate}

It is trivial that $F(\cdot)$ is monotone submodular. The two algorithms are basically different on the value of $\delta$ on step 2(a). To obtain their ratios, we observe that: at round $t$
\begin{align}
    m - F(S_t) \leq F(S^* \cup S_t) - F(S_t) \leq \sum_{e \in S^* \setminus S_t} \Delta_e F(S_t) \leq \frac{1}{\delta} OPT(V,0)  \Delta_{e_t} F(S_t) \label{equ:gua}
\end{align}
Then, their ratio is presented by Theorem. \ref{theorem:alg0_ratio}.

\begin{theorem} \label{theorem:alg0_ratio}
    Given $S_0$ that $F(S_0) \geq (1-\eta) m$; if $S$ is the returned solution, then:
    \begin{align*}
        |S \setminus S_0| \leq OPT(V,0) \frac{1}{\delta} \ln \frac{\eta m}{\alpha} + 1
    \end{align*}
\end{theorem}

\begin{proof}
    Considering after adding $e_t$, by a simple math transformation from Equ. \ref{equ:gua}, we have:
    \begin{align*}
       m - F(S_{t+1}) \leq \bigg( 1 - \frac{\delta}{OPT(V,0)} \bigg) \big(m- F(S_t)\big)
    \end{align*}
    
    Assume the algorithm terminates at $t = L$. Then at $t=L-1$, there should exist a constraint $f_i$ that $f_i(S_{L-1}) < 1-\alpha$, which means $F(S_{L-1}) < m-\alpha$. Furthermore:
    \begin{align*}
        m - F(S_{L-1}) & \leq  \bigg( 1 - \frac{\delta}{OPT(V,0)} \bigg) (m - F(S_{L-2})) \leq ... \\
        & \leq \bigg( 1 - \frac{\delta}{OPT(V,0)} \bigg)^{L-1} (m - F(S_0)) \\
        & \leq e^{-\frac{\delta (L-1)}{OPT(V,0)}} \eta m
    \end{align*}
    Thus, $L \leq OPT(V,0) \frac{1}{\delta} \ln \frac{\eta m}{\alpha} + 1$. With $S_0 = \emptyset$, the algorithm obtains the ratio of $\frac{1}{\delta}\ln \frac{m}{\alpha} + 1$.
\end{proof}
We now go over each algorithm's value of $\delta$ and their query complexity. 


With \textbf{\gr}, follow the framework, at step 2(a) \gr simply chooses $e_t = argmax_{e \in V \setminus S_t} \sum_{f_i \in \mathcal{F}_t} \Delta_e f_i(S_t)$. Then the $\delta$'s value of \gr is 1. From Theorem. \ref{theorem:alg0_ratio}, with $S_0 = \emptyset$, \gr obtains ratio of $O(\ln\frac{m}{\alpha})$. Furthermore, the algorithm scans over $V$ by at most $O(OPT(V,0) \ln \frac{m}{\alpha})$ times. Then, each $f_i$ is queried at most $O(|V|OPT\ln\frac{m}{\alpha})$ /times.


\textbf{\thr} setups a threshold $\pi$ and adds $e \in V$ to $S$ if $\sum_{i \in [m]} \Delta_e f_i(S) \geq \pi$. If no more element can be added, the algorithm reduces $\pi$ by a factor of $1-\gamma$ and scans over $V$ again. The algorithm stops when $S$ satisfies $f_i(S) \geq 1-\alpha$ for all $i \in [m]$. The pseudocode of \thr is presented by Alg. \ref{alg:threshold_gr}.

\begin{algorithm}[t]
	\caption{\thr}
	\label{alg:threshold_gr}
    \begin{flushleft}
    \textbf{Input} $V, \{f_i\}_{i \in [m]}, S^0, \gamma$ \\
	\textbf{Output} $S$ that $f_i(S) \geq 1-\alpha ~\forall~ i\in [m]$  
    \end{flushleft}
    \begin{algorithmic}[1]
        \State $t = 0$; $\pi = \max_{e \in V} \Delta_e F(S_0)$
        \State $O = $ ordered set of elements of $V$
        \State $e = $ first elements of $O$;
        \While{There exists $f_i$ that $f_i(S_t) < 1-\alpha$}
            \If{$\Delta_e F(S_t) \geq \pi$}
                \State $e_t = e$
                \State $S_{t+1} = S_t \cup \{e_t\}$;
                \State $t = t+1$
            \EndIf
            \If{$e$ is last element in $O$}
                \State $\pi = (1-\gamma) \pi$
                \State $e = $ first element in $O$
            \Else
                \State $e = $ next element in $O$
            \EndIf
        \EndWhile
    \end{algorithmic}
    \begin{flushleft}
    	\textbf{Return } $S_t$
    \end{flushleft}
\end{algorithm}

\thr always guarantees to terminate since $V$ is feasible and as long as there exists $e$ that $\sum_{i \in [m]} \Delta_e f_i(S) > 0$, $\pi$ would decrease until $e$ can be added to $S$.

In \thr, $\delta = 1-\gamma$. To show $\Delta_{e_t} F(S_t) \geq (1-\gamma) \max_{e \in V \setminus S_t} \Delta_e F(S_t)$ for each $t$, considering at the moment $e_t$ is added into $S_t$, assume $\pi$'s value is $\pi_t$, then there exists no element $e \in V \setminus S_t$ that $\Delta_e F(S_t) \geq \frac{\pi_t}{1-\gamma}$. If there exists such element, then $e$ should be added to $S_t$ when $\pi \geq \frac{\pi_t}{1-\gamma}$. The inequality follows since  $\Delta_{e_t} F(S_t) \geq \pi_t$.

In term of query complexity, we need to bound on how many times the algorithm has to scan over $V$. We have the following observation:

\begin{lemma}
If $V$ is a feasible set, given a non-feasible set $X$, there exists $e \in V$ that $\Delta_e F(X) \geq \frac{\alpha}{n}$
\end{lemma}

\begin{proof}
We use contradiction: assume there exists no such $e$. Then: 
\begin{align*}
    F(V) \leq F(X) + \sum_{e \in V \setminus X} \Delta_e F(X) < m - \alpha + n \frac{\alpha}{n} = m
\end{align*}
which contradicts to the assumption that $V$ is feasible.
\end{proof}

Therefore, \thr should terminates when $\pi \geq \frac{\alpha(1-\gamma)}{n}$. While $\pi \leq m$, the number of times the algorithm has to scan through $V$ is $\frac{1}{\gamma} \ln \frac{m n }{\alpha}$. So each $f_i$ is queried at most $O(\frac{n}{\gamma}\ln \frac{m n }{\alpha})$

\section{Tight example of \ao} 

We consider a special example of \rfd, called Robust Set Cover, defined as follows: Given a ground set $\mathcal{U}$ and a family $\mathcal{S}$ of subsets of $\mathcal{U}$, find a robust set cover $\mathcal{C} \subseteq \mathcal{S}$ of minimum size such that for all set $A \in \mathcal{S}$, $\cup_{S \in \mathcal{C}} S \setminus A = \mathcal{U}$. 

Considering the following instance of Robust Set Cover: The ground set $\mathcal{U}$ containing $n=2^k$ elements $\{e_1,...e_n\}$ and the collection $\mathcal{S}$ contains:
\begin{itemize}
    \item $S_a = S_{a^\prime} = \{e_1,e_3,...,e_{n-1}\}$ 
    \item $S_b = S_{b^\prime} = \{e_2,e_4,...,e_n\}$
    \item $S_i$ stores next $\frac{n}{2^i}$ elements to $S_{i-1}$. For example, $S_1$ contains $\{e_1,...e_{n/2}\}$, $S_2$ contains $\{e_{n/2+1}, .... e_{3n/4}\}$ and so on.
    \item $\{S_{j,i}\}_i$ are subsets of $S_j$ and $S_{j,i}$ store next $\frac{|S_j|}{2^i}$ elements to $S_{j,i-1}$.
\end{itemize}

\begin{figure}[h]
\includegraphics[width=0.8\textwidth]{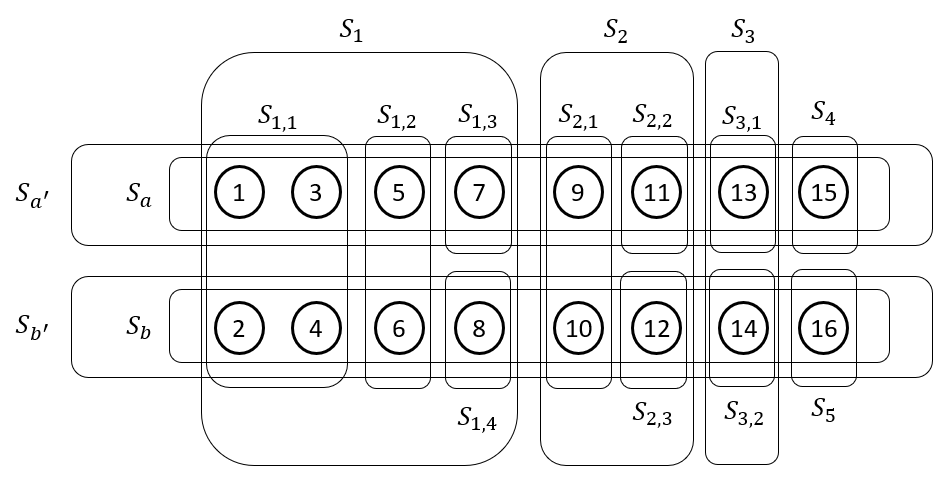}
	 	\centering
	 	\caption{Example of a tight \rfd instance with $n=16$}
	 	\label{fig:alg1_tight}
\end{figure}

Figure \ref{fig:alg1_tight} shows an example of this special instance with $n=16$.

With this instance, the optimal solution is $\mathcal{C}^{opt} = \{S_a, S_{a^\prime}, S_b, S_{b^\prime}\}$.

We make it an instance of \rfd by defining an input set of submodular functions $\{f_e\}_{e \in \mathcal{U}}$, where $f_e(\mathcal{C})$ is the number of sets in $\mathcal{C}$ containing $e$. The threshold $T_e = 1$ for all $e \in \mathcal{U}$. With this \rfd instance, to find a feasible set cover, we simply set $\alpha$ large and close to $1$. 

So, if applying \ao to this \rfd instance, $\mathcal{S}_1$ returned from \az in line \ref{line:first_out} of Alg. \ref{alg:alg_1} may contain all $S_i$ sets. It is trivial that $|\mathcal{S}_1| \leq O(\ln n) |\mathcal{S}^{opt}|$. Also, any removal of a set $S_i$ from $\mathcal{S}_1$ makes it violate at least one constraint since $S_i$s are disjoint.

Then, again, applying \az to $\mathcal{S} \setminus S_i$ for each $S_i$ may result to all $S_{i,j}$s set are added. Then the final solution's size would be $O(\sum_{S_i \in \mathcal{S}_1} \ln |S_i| ) \leq O(|\mathcal{S}_1| \ln n)$, which is tight to our analysis.

\end{document}